\pdfoutput=1
\documentclass[11pt]{article}
\usepackage[letterpaper,margin=1in]{geometry}
\usepackage[T1]{fontenc}
\usepackage{lmodern}
\usepackage{microtype}
\usepackage[font=small]{caption}
\usepackage{cite}
\usepackage{amsmath,amssymb,amsfonts}
\usepackage{graphicx}
\graphicspath{{images/}}
\usepackage{textcomp}
\usepackage{booktabs}
\usepackage{multirow}
\usepackage[hidelinks]{hyperref}

\begin{document}

\title{Sub-Synchronous Control Interaction Risk of Large Data Center Loads: A Frequency-Scan Assessment with Cross-Model Validation}

\author{Salish Maharjan$^{1}$, Junyuan Zheng$^{2}$, Lung-An Lee$^{1}$, Min Lwin$^{1}$, Zhaoyu Wang$^{2}$\\[6pt]
{\normalsize $^{1}$INS Engineering, TX, USA \quad $^{2}$Iowa State University, IA, USA}\\
{\normalsize smaharjan, vlee, mlwin@ins-engineering.com; zhengjy, wzy@iastate.edu}}
\date{\vspace{-4pt}\fbox{\parbox{0.8\textwidth}{\small\centering This work has been submitted to the IEEE for possible publication. Copyright may be transferred without notice, after which this version may no longer be accessible.}}}

\maketitle

\begin{abstract}
Data centers are now one of the fastest-growing loads on the bulk power system, and most of their demand reaches the grid through power converters inside uninterruptible power supplies (UPS). Whether these loads can take part in sub-synchronous control interaction (SSCI), as wind and HVDC converters do, has not been studied extensively. We examine this question with impedance-based frequency scans at the point of common coupling (PCC). Using a detailed proprietary model (INS-ISU), we scan the motor, variable-frequency-drive (VFD), lighting, and UPS-fed information-technology (IT) branches of a representative facility separately. Only the UPS in double-conversion mode shows negative resistance; it covers the whole sub-synchronous band and continues up to about 170~Hz. The other branches, and the UPS in bypass mode, stay passive. We then repeat the scans on three public models from PNNL, TAMU, and NLR. None of them shows negative resistance in bypass mode, while three of the four models do in double-conversion mode. Lowering the load from rated to light load makes the negative resistance deeper, as expected for a constant-power load, so a lightly loaded facility in double-conversion mode is the worst case. Raising the UPS DC-link capacitance pushes the negative resistance into a narrow band around the fundamental. These findings give a basis for screening data centers for SSCI during interconnection studies.
\end{abstract}


\section{Introduction}
\label{sec:intro}

Data center electricity use is rising quickly. In the United States, it more than doubled between 2017 and 2023, and the latest national estimate puts it at 6.7--12\% of total consumption by 2028~\cite{shehabi2024lbnl}. The IEA expects global demand from data centers to roughly double again, to about 945~TWh, by 2030~\cite{iea2025energyai}. Single campuses of several hundred megawatts are no longer unusual, and many of them connect directly at transmission voltage. What sets these loads apart from older industrial plants is how they draw power. Servers sit behind UPS units, and in a double-conversion UPS every watt passes through a rectifier, a DC bus, and an inverter before it reaches the IT equipment.

Grid operators have already seen what large data centers can do during disturbances. In July 2024, a single 230-kV line fault in the Eastern Interconnection led to the loss of roughly 1,500~MW of data center load~\cite{nerc2025incident}. NERC has since listed stability among the main risks of emerging large loads~\cite{nerc2025largeloads}, and AI training clusters are known to swing their demand by tens or even hundreds of megawatts, enough to excite grid resonances~\cite{choukse2025power}. Most of this work deals with ride-through, load loss, or forced oscillations. The small-signal interaction between the converter controls inside a data center and the transmission network has received little attention so far.

For generation, this kind of interaction is well known. In 2009, Type-3 wind turbines in ERCOT interacted with a series-compensated line and produced growing sub-synchronous oscillations~\cite{adams2012ercot}. Field records from China show similar behavior in wind farms on series-compensated lines~\cite{xie2017characteristic} and in direct-drive wind farms on weak grids~\cite{liu2017subsynchronous}, and an IEEE PES task force has collected these events in~\cite{cheng2020windsso}. VSC-HVDC terminals face related resonance problems~\cite{cigre2019tb754}. In each case, the converter controls act like a negative resistance over part of the sub-synchronous range. If that resistance outweighs the network damping at a resonant frequency, the oscillation grows.

Impedance-based methods are a common way to study these problems. The idea goes back to Middlebrook's input-filter criterion~\cite{middlebrook1976}: stability follows from the ratio of the source and load impedances~\cite{sun2011impedance}. Converter impedances can be derived in the sequence domain~\cite{cespedes2014impedance} or in the d-q frame~\cite{wen2016dq}, and a converter cannot destabilize resonances that fall where its admittance is passive, that is, where its real part is positive~\cite{harnefors2016passivity,wang2019harmonic}. When the controller details are hidden, the impedance can still be measured by frequency scans of an electromagnetic transient (EMT) model~\cite{ren2016refined,cheng2013reactance}. This is the usual situation in interconnection studies, where the vendor supplies a black-box model.

There are good reasons to expect negative resistance from data center loads. A converter that holds its power constant presents a negative incremental resistance inside its control bandwidth~\cite{emadi2006constant}, and the synchronization loop of a grid-tied converter can add a similar effect~\cite{wen2016dq}. A double-conversion UPS regulates its DC bus and feeds an inverter that serves the IT load~\cite{bekiarov2002ups}, so it has both features. EMT models of data centers have recently become available~\cite{ross2025pnnl,wang2026tamu,shah2026nlr}, but they have mostly been used for ride-through and load-loss studies. We are not aware of any work that evaluates whether data center loads pose an SSCI risk, which components within the facility contribute most, or how the UPS operating mode, loading, and design impact the impedance characteristics.

This paper makes four contributions:

\begin{enumerate}
\item We scan each branch of a representative data center (motor, VFD, lighting, and UPS-fed IT load) in a detailed proprietary model (INS-ISU) and show that only the double-conversion UPS presents negative resistance.
\item We repeat the scans on three independently built public models (PNNL, TAMU, and NLR) with the same procedure, so the result does not rest on a single model.
\item We vary the UPS operating mode (bypass and double conversion) and the loading (100\%, 50\%, and 10\%) and find that a lightly loaded double-conversion UPS is the worst case in terms of negative resistance presented to the grid.
\item We show that a larger UPS DC-link capacitance confines the negative resistance to a narrow band around the fundamental, and we outline how data centers can be screened in interconnection studies.
\end{enumerate}

Section~\ref{sec:model} describes the load composition and the scan procedure. Section~\ref{sec:evidence} gives the branch-level results for the INS-ISU model, and Section~\ref{sec:crossval} compares them with the open-source models. Section~\ref{sec:sensitivity} covers loading and DC-link capacitance. Section~\ref{sec:discussion} summarizes the results and their implications, and Section~\ref{sec:conclusion} concludes the paper.

\section{Data Center Load Composition and Frequency Scan Methodology}
\label{sec:model}

\subsection{Load Composition}
\label{sec:composition}

\begin{figure}[!htbp]
  \centering
  \includegraphics[width=0.6\linewidth]{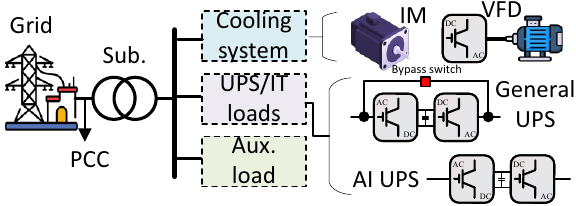}
  \caption{General composition of data center loads.}
  \label{dc_config}
\end{figure}

Fig.~\ref{dc_config} shows the general layout of a data center load. The components within the facility are grouped into three categories.

\begin{itemize}
\item \textbf{Motor loads (15\%)} are the induction motors of the chillers and fans used for cooling. They are fed either directly from the grid or through VFDs.
\item \textbf{IT loads (80\%)} are servers and network equipment, supplied through power distribution units (PDUs) behind UPS systems. Many UPS units run in bypass mode most of the time and switch to double-conversion mode (DCM) during voltage disturbances so that the IT load is not interrupted. Some UPS units in AI training facilities have no bypass path and run in DCM continuously.
\item \textbf{Lighting loads (5\%)} are mostly LED fixtures with electronic drivers, which behave as constant-power loads.
\end{itemize}

\subsection{Frequency Scan Methodology}
\label{sec:methodology}

Similar to the typical procedure described in \cite{{xie2017characteristic,ren2016refined}}, we apply the same frquency scan to every model, proprietary or open-source, so that the results can be compared directly. The steps are as follows.

\begin{enumerate}
\item A small series voltage perturbation $\Delta V(f)$ is added to the grid voltage at the PCC. Its frequency $f$ is swept from 5 to 1000~Hz in steps of 1 to 10~Hz.
\item The current is recorded with and without the perturbation. The difference is passed through a discrete Fourier transform, and its component at $f$ gives $\Delta I(f)$. The positive-sequence impedance is then
\begin{equation}
Z(f) = \frac{\Delta V(f)}{\Delta I(f)} = R(f) + jX(f).
\label{eq:z}
\end{equation}
\item We plot the resistance $R(f)$ and reactance $X(f)$ against frequency. Wherever $R(f)$ is negative, the load can sustain an oscillation at that frequency instead of damping it, which flags a possible SSCI risk.
\end{enumerate}

\subsubsection{Stability Interpretation}
Let $Z_g(f) = R_g(f) + jX_g(f)$ be the network impedance seen from the PCC. The network and the load form a series resonance at the frequency $f_r$ where their reactances cancel,
\begin{equation}
X_g(f_r) + X(f_r) = 0.
\label{eq:resonance}
\end{equation}
The oscillation at $f_r$ decays only if the total resistance is positive~\cite{cheng2013reactance,harnefors2016passivity}:
\begin{equation}
R_g(f_r) + R(f_r) > 0.
\label{eq:damping}
\end{equation}
For a series-compensated line, $f_r \approx f_0\sqrt{X_C/X_L}$, where $f_0$ is the fundamental frequency and $X_C$ and $X_L$ are the capacitive and total inductive reactances at $f_0$. Practical compensation levels place $f_r$ in the sub-synchronous range. Transmission networks have little resistance, so a load with negative $R(f)$ in this range can violate (\ref{eq:damping}). A load whose resistance is positive everywhere only adds damping.

\subsubsection{Negative Resistance of Constant-Power Loads}
Suppose a converter holds its power at $P$ within its control bandwidth. At voltage $V$ it draws $I = P/V$, so its incremental resistance is
\begin{equation}
r = \frac{\partial V}{\partial I} = -\frac{V^2}{P}
\label{eq:cpl}
\end{equation}
\cite{emadi2006constant}. The resistance is negative, and its magnitude grows as $P$ falls. Above the control bandwidth, the controller can no longer hold the power constant, and the resistance turns positive again, set by the passive components. We therefore expect a tightly regulated double-conversion UPS is therefore expected to show negative resistance around the fundamental that extends into the sub-synchronous range by an amount set by its control bandwidth, with a magnitude that is largest at light load.

\subsubsection{Excluded Band Around the Fundamental}
\label{sec:excluded}
The scans leave out the band between about 45 and 65~Hz. We exclude it for three reasons. First, the measurement is unreliable there. The response at the perturbation frequency $f$ has to be separated from a fundamental current that is far larger. As $f$ approaches $f_0$, the two components can be resolved only with a long DFT window, and any small mismatch in the fundamental current between the runs with and without perturbation then dominates $\Delta I(f)$ and corrupts $Z(f)$. Second, a perturbation at $f$ also produces a response at the mirror frequency $2f_0 - f$ through the phase-locked loop and other asymmetric controls~\cite{wen2016dq}. Away from $f_0$ this coupling is weak, and the positive-sequence impedance in (\ref{eq:z}) describes the load well. Near $f_0$, the two frequencies merge, and a single sequence impedance is no longer an adequate description. Third, this band is not where SSCI normally arises. For typical series-compensation levels of 25--50\%, the electrical resonance $f_r \approx f_0\sqrt{X_C/X_L}$ lies at about 30--42~Hz, well inside the scanned range. Oscillations within a few hertz of $f_0$ appear instead as slow modulations of power and voltage magnitude, and they are usually studied with electromechanical (RMS) models together with the voltage and frequency controls of the plant, not with EMT frequency scans. We therefore leave this band out of the screening. Where a load shows negative resistance right up to the edges of the band, as the double-conversion UPS does, the band should be examined in the combined load--network study discussed in Section~\ref{sec:discussion}.

The procedure needs no knowledge of the controller structure or its parameters. This is what lets us use it on a proprietary model and on open-source models alike, and it works the same way on the black-box models that original equipment manufacturers (OEMs) supply for interconnection studies.

\section{Evidence from the Proprietary INS-ISU Model}
\label{sec:evidence}

\subsection{Model Disclosure Statement}

The INS-ISU model is a detailed proprietary data center load model~\cite{zheng2026emt}. Because of confidentiality, we disclose only its high-level structure (branch composition, interconnection topology, and modeling assumptions) and the aggregate scan results needed to support our conclusions. Internal control parameters, component models, and tuning data are withheld.

\subsection{Branch-Isolated Frequency Scan Results}

\begin{figure}[!htbp]
  \centering
\begin{minipage}[t]{0.485\linewidth}
  \centering
  \includegraphics[width=\linewidth]{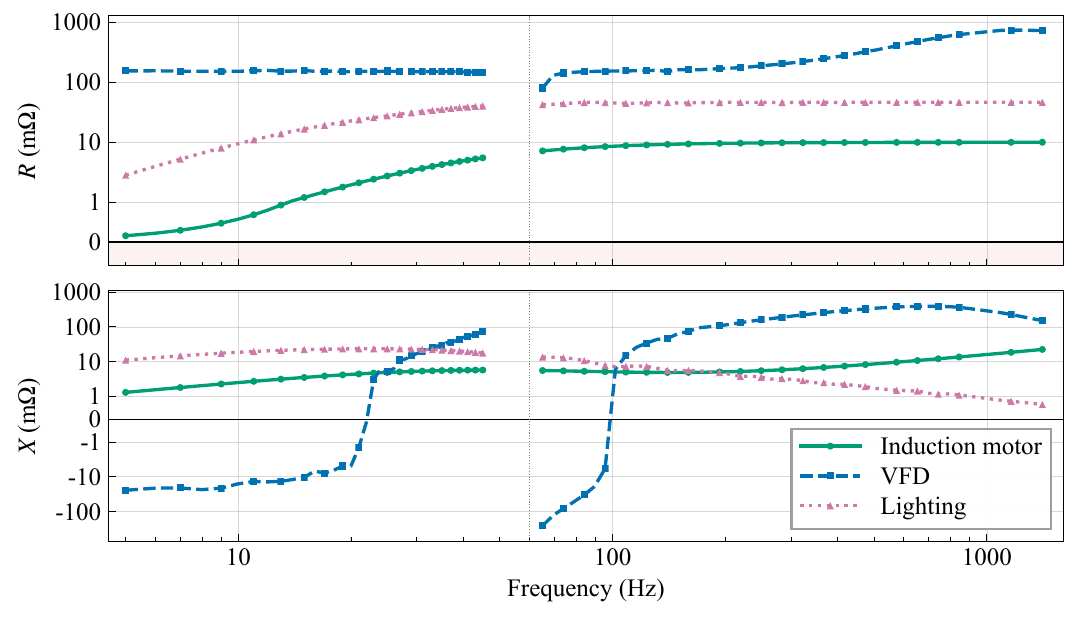}
  \caption{Positive-sequence impedance of the induction motor, VFD, and lighting}
  \label{isu_im_vfd_lighting}
\end{minipage}\hfill
\begin{minipage}[t]{0.485\linewidth}
  \centering
  \includegraphics[width=\linewidth]{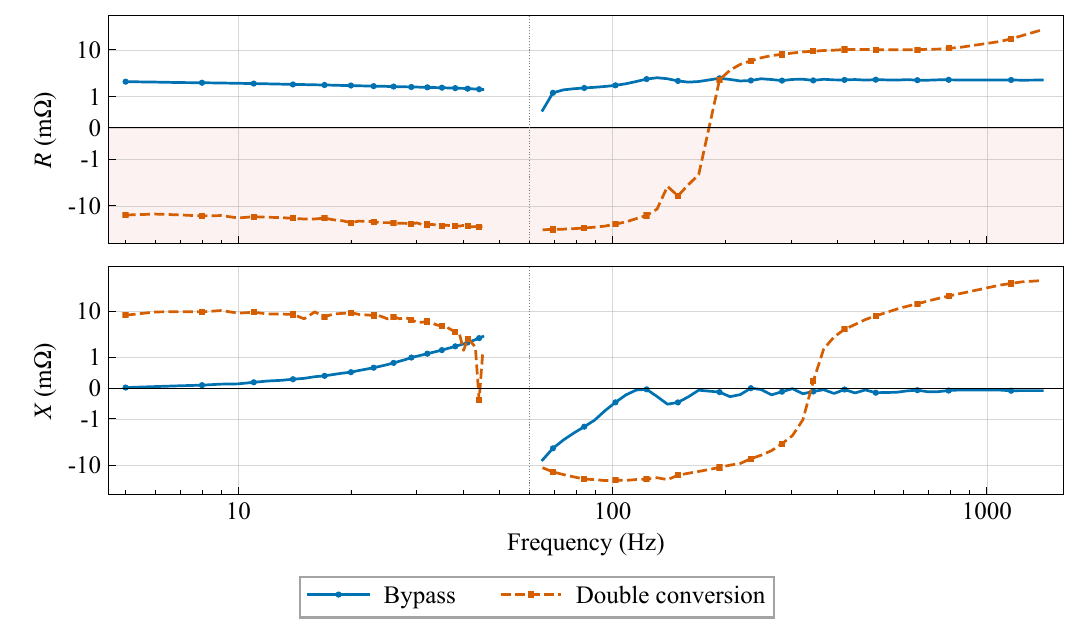}
  \caption{Positive-sequence impedance of the INS-ISU UPS branch(nominal 100~MW).}
  \label{isu_ups_dcm}
\end{minipage}
\end{figure}

We first scanned each branch on its own to see which ones could contribute to SSCI. Fig.~\ref{isu_im_vfd_lighting} shows $R(f)$ and $X(f)$ for the induction motor, VFD, and lighting branches; the band around the fundamental is left out of the scan for the reasons given in Section~\ref{sec:excluded}. None of the three branches enters the negative-resistance region at any frequency. The VFD has the largest resistance, about 150~m$\Omega$ across the sub-synchronous range, rising to about 700~m$\Omega$ near 1~kHz. The lighting branch rises from about 3~m$\Omega$ at 5~Hz to about 40~m$\Omega$ near the fundamental and stays almost flat above it. The induction motor gives the least damping. Its resistance drops to a fraction of a milliohm at the lowest frequencies and approaches 10~m$\Omega$ in the super-synchronous range, but it never goes negative. The reactances differ more. The motor and lighting branches are inductive throughout, while the VFD is capacitive at low frequencies and crosses zero near 20~Hz below the fundamental and near 100~Hz above it. The sign of the reactance does not matter here: with $R(f) > 0$, all three branches damp oscillations rather than sustain them. The UPS branch behaves differently, as shown next.

Fig.~\ref{isu_ups_dcm} compares the 100-MW UPS branch in bypass and double-conversion modes. The operating mode decides whether the UPS adds or removes damping. In bypass mode, the resistance stays positive at about 1--3~m$\Omega$ over the whole range, and the reactance is small, slightly inductive below the fundamental and close to zero above about 100~Hz. In double-conversion mode, the resistance is negative, at about $-10$ to $-15$~m$\Omega$, from 5~Hz up to the fundamental. It stays negative above the fundamental up to roughly 170~Hz, then turns positive and reaches several tens of milliohms near 1~kHz. This is the constant-power behavior predicted by (\ref{eq:cpl}): the tightly regulated rectifier--inverter stage looks like a negative resistance within its control bandwidth. The reactance in this mode is inductive at low frequencies, capacitive just above the fundamental, and crosses zero again near 300~Hz. Because the negative resistance covers the entire sub-synchronous band, where series-compensated lines resonate, the double-conversion UPS is the main source of SSCI risk in the facility. It also carries about 80\% of the load, so the risk depends heavily on how much UPS capacity runs in double-conversion mode. We return to this point in Section~\ref{sec:sensitivity}.

\section{Cross-Validation with Open-Source Data Center Models}
\label{sec:crossval}

\subsection{Open-Source Reference Models}
To compare the results of a INS-ISU model of Section~\ref{sec:evidence}, the same scan procedure (Section~\ref{sec:methodology}) is applied to three publicly available models developed by other groups:

\begin{itemize}
\item \textbf{PNNL model}: the generic data center model library (DML) built in PSCAD by Pacific Northwest National Laboratory. It aggregates double-conversion UPS units and single-phase power factor correction converters for grid-level EMT studies~\cite{ross2025pnnl}.
\item \textbf{TAMU model}: a modular PSCAD model of a double-conversion-UPS-based AI data center from Texas A\&M University and ERCOT. It includes the rectifier, inverter, battery DC/DC converter, VFD cooling, and aggregated computing loads, with bypass, normal, and battery-backup modes in both switching and average form~\cite{wang2026tamu}.
\item \textbf{NLR model}: the PSCAD EMT library of the National Laboratory of the Rockies (NLR, formerly NREL). It contains an IT rack load with switching or averaged power supply units, a generic UPS, and a motor representing the cooling load~\cite{shah2026nlr}.
\end{itemize}

\begin{figure}[!htbp]
  \centering
\begin{minipage}[t]{0.485\linewidth}
  \centering
  \includegraphics[width=\linewidth]{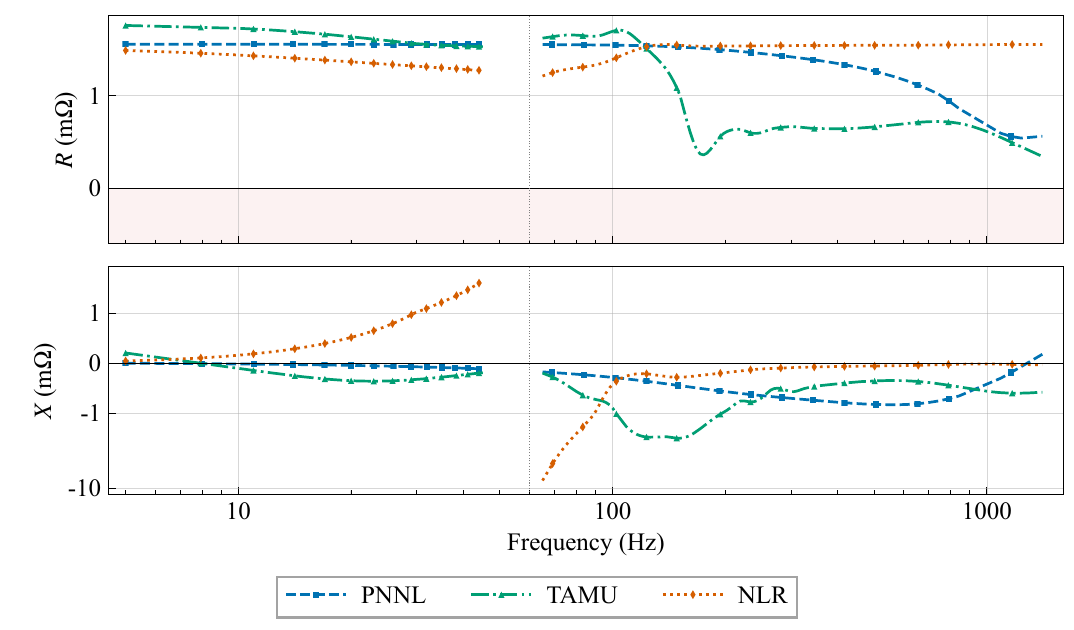}
  \caption{Positive-sequence impedance of the opensource models in bypass mode at 100~MW. The shaded area marks negative resistance.}
  \label{pnnl_bypass_100}
\end{minipage}\hfill
\begin{minipage}[t]{0.485\linewidth}
  \centering
  \includegraphics[width=\linewidth]{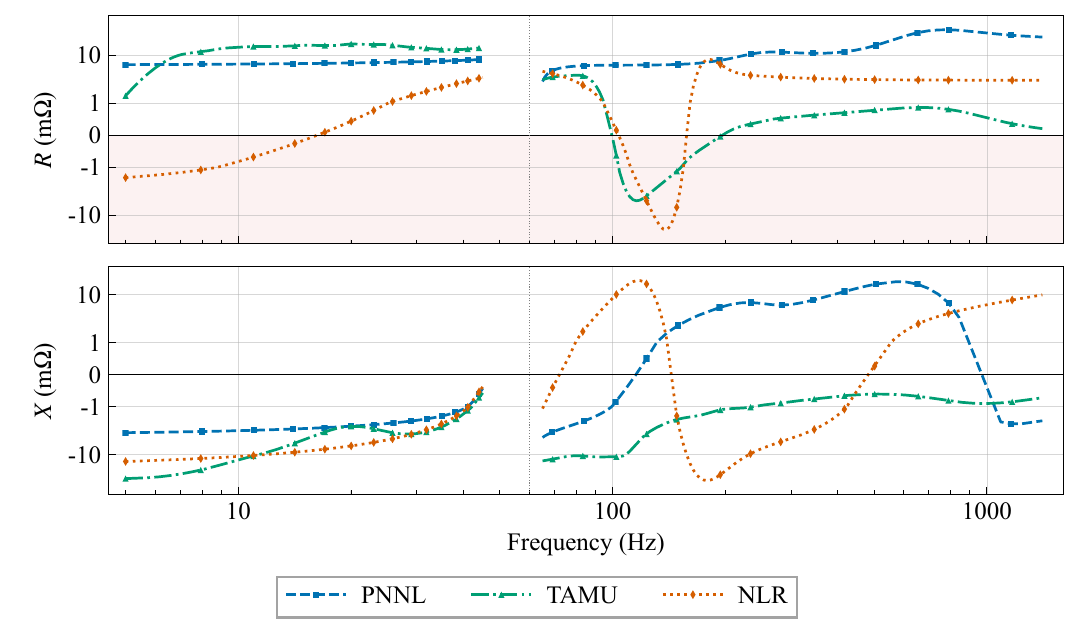}
  \caption{Positive-sequence impedance of the opensource models in double-conversion mode at 100~MW.}
  \label{pnnl_dcm_100}
\end{minipage}
\end{figure}

\subsection{Comparison of Bypass UPS}

Fig.~\ref{pnnl_bypass_100} shows the three open-source models with the UPS in bypass mode at 100~MW. All three keep a positive resistance over the full range. Below the fundamental, the resistance is nearly flat, between about 1.3 and 1.8~m$\Omega$. Above it, the models part ways: the NLR model stays near 1.5~m$\Omega$, the PNNL model falls slowly to about 0.5~m$\Omega$ near 1.5~kHz, and the TAMU model drops to about 0.4~m$\Omega$ near 170~Hz and stays below 1~m$\Omega$ from there on. The reactance is small in every case. The NLR model is inductive below the fundamental, reaching about 1.5~m$\Omega$ at 45~Hz, while the PNNL and TAMU models are slightly capacitive. Above the fundamental, all three are mildly capacitive over most of the range. These curves match the bypass result of the INS-ISU model in Fig.~\ref{isu_ups_dcm}, which also stays at about 1--3~m$\Omega$. All four models therefore agree that a data center with its UPS in bypass does not introduce negative damping and poses no SSCI risk.

\subsection{Comparison of Double-Conversion UPS}

Fig.~\ref{pnnl_dcm_100} repeats the comparison in double-conversion mode at 100~MW. Here the models disagree noticeably, and two of them go negative. The NLR model has negative resistance at low frequencies, about $-1.5$~m$\Omega$ at 5~Hz, crossing zero near 17~Hz and rising to about 3~m$\Omega$ at 45~Hz. It also has a deep negative band above the fundamental, between about 100 and 160~Hz, reaching about $-20$~m$\Omega$ near 140~Hz. The TAMU model is positive below the fundamental, at about 10--15~m$\Omega$ above 10~Hz, but shows a similar negative band between about 100 and 150~Hz, with a minimum of about $-6$~m$\Omega$ near 120~Hz. The PNNL model stays positive everywhere, at about 6~m$\Omega$ below the fundamental, rising to about 20~m$\Omega$ near 800~Hz. Below the fundamental, all three models are capacitive, with reactance between about $-1$ and $-20$~m$\Omega$, and the NLR model shows a sharp reactance resonance near 120--190~Hz that lines up with its negative band. The INS-ISU model in Fig.~\ref{isu_ups_dcm} is negative across the sub-synchronous range and up to about 170~Hz. The NLR model reproduces its low-frequency negative resistance, and both the NLR and TAMU models reproduce its negative band below about 170~Hz. The PNNL model has none, which shows that the size and frequency range of the negative damping depend on how each converter is controlled and tuned. Still, double-conversion mode is the only condition in which any model goes negative, so this is the mode that needs SSCI screening.

\section{Sensitivity and Mitigation Studies on the INS-ISU Model}
\label{sec:sensitivity}

\subsection{Sensitivity to Loading Levels in Bypass System}

Fig.~\ref{isu_bypass_load} shows the INS-ISU model in bypass mode at 100, 50, and 10~MW. The resistance stays positive at every loading level, so a lighter load does not push the bypass system into the negative region. In fact, the resistance rises as the load falls, roughly in inverse proportion. Below the fundamental, it is about 1.4--2.2~m$\Omega$ at 100~MW, 3--4~m$\Omega$ at 50~MW, and 15--22~m$\Omega$ at 10~MW, dropping slightly toward the fundamental in each case. Above the fundamental, it settles at about 2, 4.5, and 8--16~m$\Omega$, respectively. The smallest margin is at 100~MW just above the fundamental, where the resistance dips to about 0.5~m$\Omega$ near 65~Hz before recovering by 100~Hz. The 10-MW case has a short dip to about 3~m$\Omega$ near 140~Hz, still well above zero. The reactance is inductive below the fundamental at all three levels and grows toward 45~Hz, reaching about 2.5, 6, and 20~m$\Omega$ at 100, 50, and 10~MW. Just above the fundamental it is capacitive. It returns to nearly zero by about 110~Hz at 100 and 50~MW, whereas at 10~MW it starts near $-80$~m$\Omega$ and crosses zero only near 350~Hz. In short, the bypass system damps oscillations at any loading, and a lightly loaded data center in bypass mode do no indicate any SSCI risk.

\subsection{Sensitivity to Loading Levels in Double-Conversion System}

\begin{figure}[!htbp]
  \centering
\begin{minipage}[t]{0.485\linewidth}
  \centering
  \includegraphics[width=\linewidth]{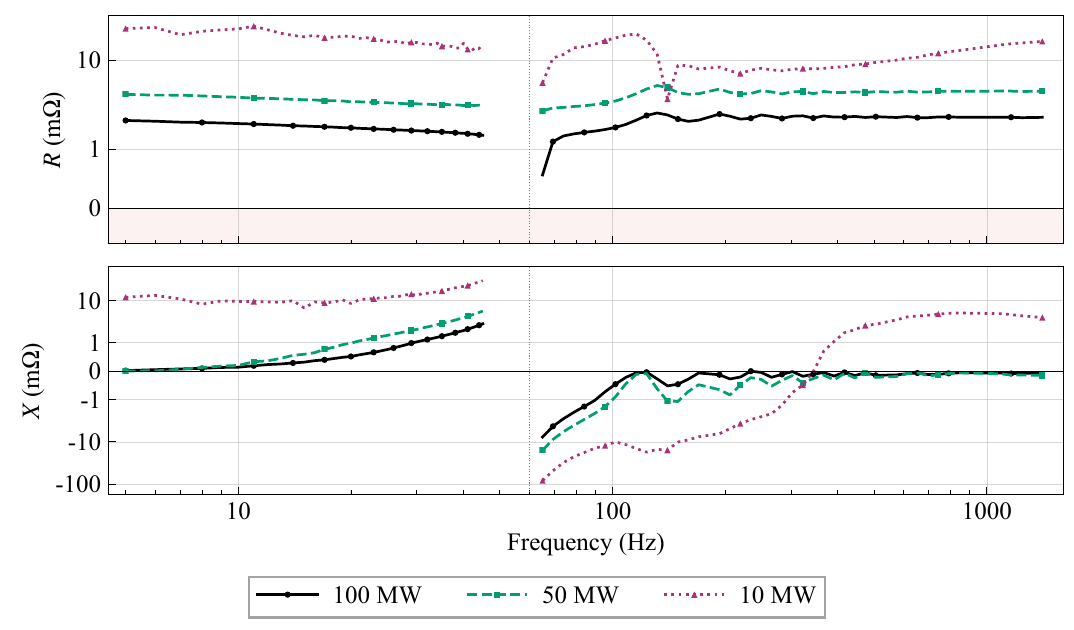}
  \caption{Positive-sequence impedance of the INS-ISU model in bypass mode at three loading levels. The shaded area marks negative resistance.}
  \label{isu_bypass_load}
\end{minipage}\hfill
\begin{minipage}[t]{0.485\linewidth}
  \centering
  \includegraphics[width=\linewidth]{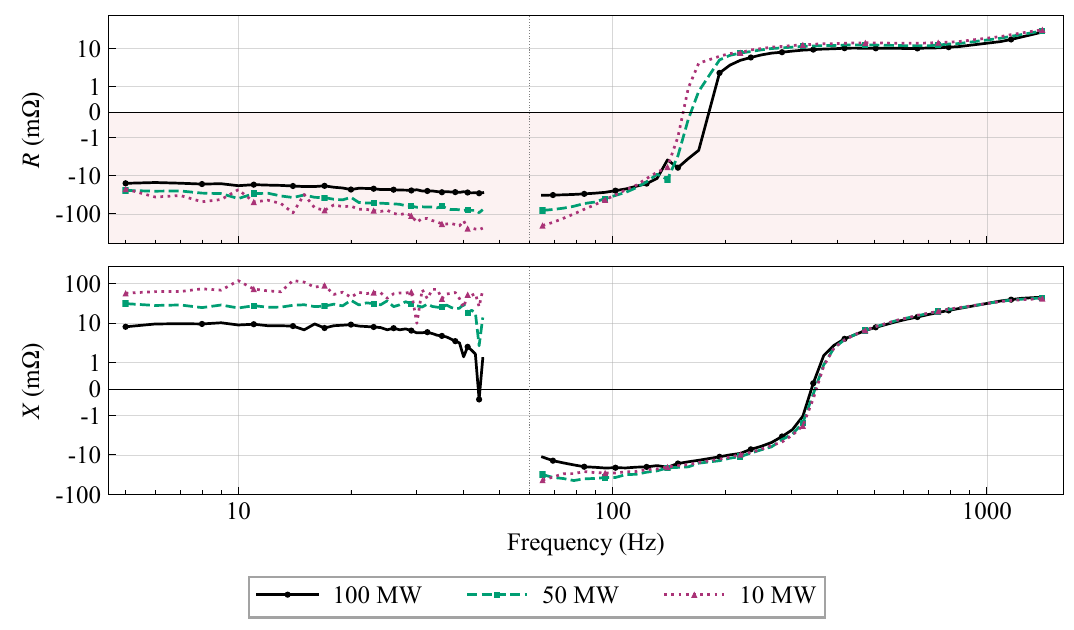}
  \caption{Positive-sequence impedance of the INS-ISU model in double-conversion mode at three loading levels. The shaded area marks negative resistance.}
  \label{isu_dcm_load}
\end{minipage}
\end{figure}

Fig.~\ref{isu_dcm_load} repeats the loading study in double-conversion mode. At all three loading levels, the resistance is negative across the whole sub-synchronous band and stays negative above the fundamental up to about 150--190~Hz. A lighter load makes it worse. Near 40~Hz, the resistance is about $-20$~m$\Omega$ at 100~MW, $-70$~m$\Omega$ at 50~MW, and $-150$~m$\Omega$ at 10~MW. This follows the incremental resistance $-V^2/P$ of a constant-power load in (\ref{eq:cpl}), whose magnitude grows as $P$ shrinks. The sub-synchronous reactance is inductive and also grows at light load, while all three cases change from capacitive to inductive near 350~Hz. According to (\ref{eq:damping}), the damping of a sub-synchronous mode depends on the sum of the network and load resistances at the resonant frequency, and the small network resistance has a harder time offsetting a larger negative load resistance. A lightly loaded data center in double-conversion mode therefore presents the largest negative resistance to the grid.

\subsection{Avoiding Negative Resistance in Double-Conversion System}

In bypass mode, the UPS passes utility power to the IT load with little power-electronic involvement, so the impedance seen by the grid is close to that of the IT/PDU load itself. In double-conversion mode, the UPS rectifies the incoming AC to a DC bus and inverts it again for the IT load. This decouples the load from the grid at the fundamental frequency, but it also places the rectifier and inverter control loops (current control, DC-bus voltage regulation, and the phase-locked loop)~\cite{bekiarov2002ups} in the grid-facing impedance. These loops have limited bandwidth and depend on loading, which is consistent with the negative resistance found in Sections~\ref{sec:evidence}--\ref{sec:crossval} and with its growth at light load.

\begin{figure}[!htbp]
  \centering
\begin{minipage}[t]{0.485\linewidth}
  \centering
  \includegraphics[width=\linewidth]{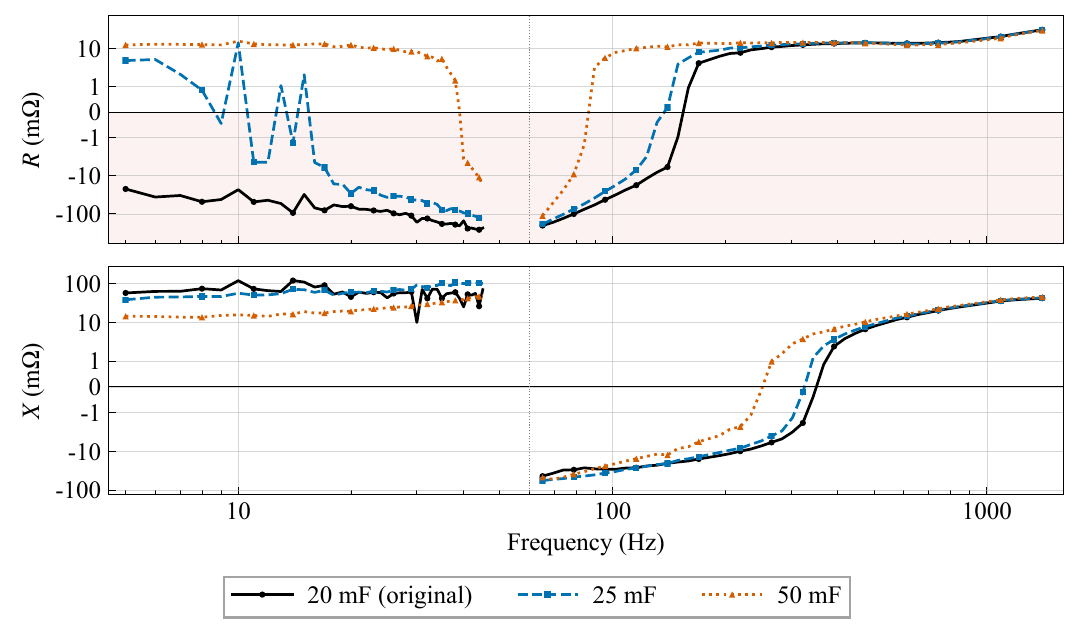}
  \caption{Positive-sequence impedance of the INS-ISU model in double-conversion mode at a nominal load of 10~MW for three DC-link capacitors of the UPS.}
  \label{isu_dcm_cap_10}
\end{minipage}\hfill
\begin{minipage}[t]{0.485\linewidth}
  \centering
  \includegraphics[width=\linewidth]{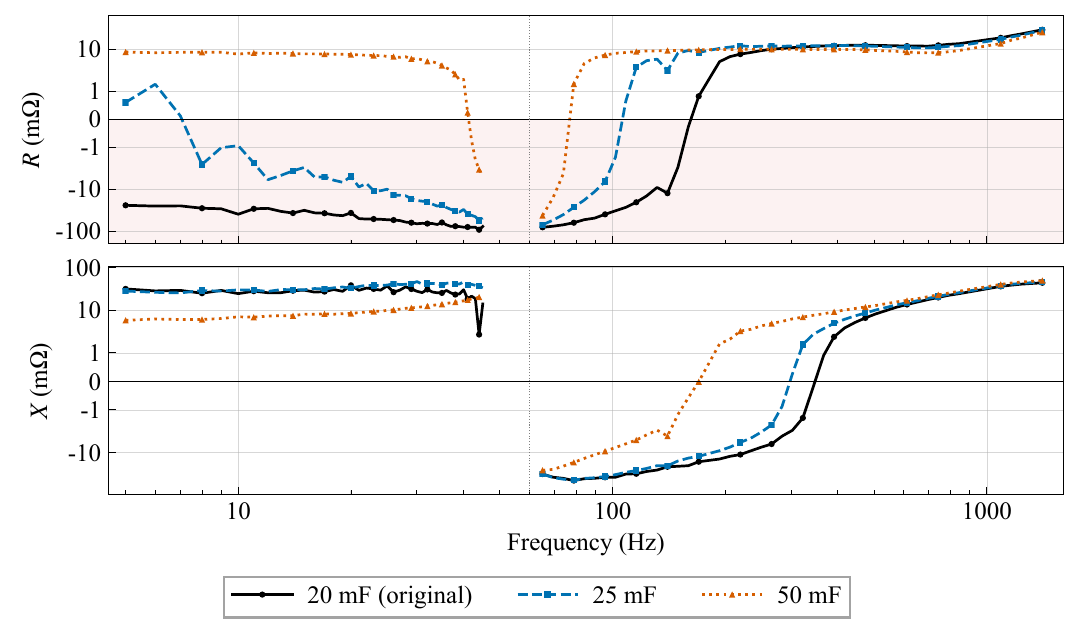}
  \caption{Positive-sequence impedance of the INS-ISU model in double-conversion mode at a nominal load of 50~MW for three DC-link capacitors of the UPS.}
  \label{isu_dcm_cap_50}
\end{minipage}
\end{figure}

\begin{figure}[!htbp]
  \centering
  \includegraphics[width=0.485\linewidth]{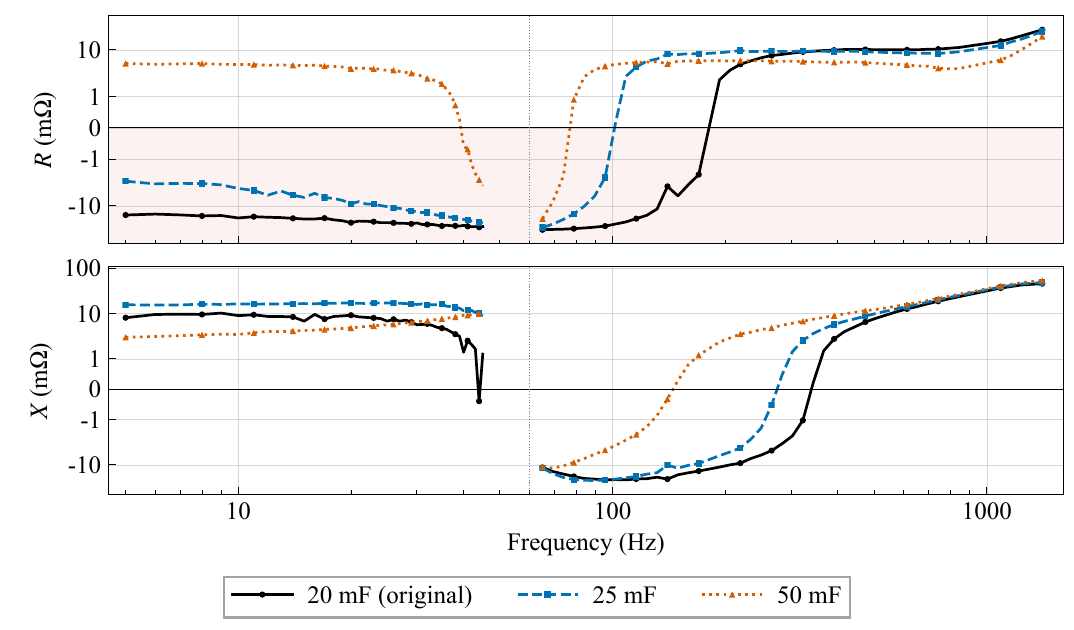}
  \caption{Positive-sequence impedance of the INS-ISU model in double-conversion mode at a nominal load of 100~MW for three DC-link capacitors of the UPS.}
  \label{isu_dcm_cap_100}
\end{figure}

The dynamics of the DC-bus voltage loop are set jointly by the controller gains and the DC-link capacitance. Here we vary the capacitance with the gains held fixed, which lowers the loop bandwidth; re-tuning the gains would be an alternative way to achieve the same effect. Figs.~\ref{isu_dcm_cap_10}--\ref{isu_dcm_cap_100} compare the original 20~mF with 25 and 50~mF at 10, 50, and 100~MW. With 20~mF, the resistance is negative across the whole sub-synchronous band and up to about 160--185~Hz at every loading level. Going to 25~mF narrows the negative band above the fundamental to below about 105--140~Hz and reduces the low-frequency negative resistance. It restores positive resistance only below about 8~Hz at 10 and 50~MW, though, and nowhere in the sub-synchronous band at 100~MW. With 50~mF, the resistance stays positive at about 5--12~m$\Omega$ up to about 35--40~Hz and again above about 75--90~Hz at all three loading levels, leaving only a narrow negative band around the fundamental. The larger capacitance also lowers the sub-synchronous reactance and moves the capacitive-to-inductive crossover from about 350~Hz down to about 150--260~Hz; all curves converge above about 200~Hz.This substantially narrows the frequency range over which the UPS could contribute to SSCI. A comparable effect would be expected from retuning the controller. The band close to the fundamental remains negative, however, and since it lies at the edge of the excluded band (Section~\ref{sec:excluded}), it should still be checked in interconnection studies.

\section{Results Summary and Discussion}
\label{sec:discussion}

Table~\ref{tab:summary} collects the results of all 24 scan cases: four models, two UPS modes, and three loading levels (100, 50, and 10~MW). Bypass operation never produced negative resistance, in any model or at any load. Negative resistance appeared only in double-conversion mode, in three of the four models, and at different frequencies in each. Where it appeared, a lighter load usually made it deeper or wider, as (\ref{eq:cpl}) predicts for a constant-power load. A lightly loaded data center in double-conversion mode is therefore the worst case for SSCI, and a substantially larger DC-link capacitance removes most of the negative damping from the sub-synchronous band.

\begin{table}[!htbp]
\caption{Summary of Negative-Resistance Regions Across Models}
\label{tab:summary}
\centering
\small
\setlength{\tabcolsep}{8pt}
\begin{tabular}{llll}
\toprule
Model & UPS mode & Negative-$R$ band & Min.\ $R$ (m$\Omega$) at \\
 & & & 100/50/10~MW \\
\midrule
\multirow{2}{*}{INS-ISU} & Bypass & None & $>0$ \\
 & DCM & 5 to 150--190~Hz & $-20$/$-70$/$-150$\textsuperscript{a} \\
\midrule
\multirow{2}{*}{PNNL} & Bypass & None & $>0$ \\
 & DCM & None & $>0$ \\
\midrule
\multirow{2}{*}{TAMU} & Bypass & None & $>0$ \\
 & DCM & 100--190~Hz & $-5$/$-8$/$-25$ \\
\midrule
\multirow{2}{*}{NLR} & Bypass & None & $>0$ \\
 & DCM & $<$17--33~Hz; 90--160~Hz & $-20$/$-100$/$-150$ \\
\bottomrule
\multicolumn{4}{l}{\textsuperscript{a}Evaluated near 40~Hz.}
\end{tabular}
\end{table}

For SSCI screening of a large data center, these results point to the UPS-fed IT branch. The models also differ a great deal, from no negative resistance in the PNNL model to a broad negative band in the INS-ISU model, because the negative damping depends on converter control design and tuning. Generic models can show whether a risk may exist, but confirming and quantifying it requires a scan of the vendor's own model.

A UPS that shows positive resistance over the scanned range effectively poses no SSCI risk once connected. One that shows negative resistance, has the potential for control interaction with the network or nearby IBRs, and should be studied together with the grid: a combined frequency scan of the data center and the network~\cite{cheng2013reactance} shows whether the network resistance can offset the negative load resistance at the resonant frequencies. The answer depends on the grid model, which remains uncertain until the point of interconnection (POI) is fixed. The load-side scans in this paper should therefore be read as a screening step, not a final assessment.

\section{Conclusion}
\label{sec:conclusion}

We used impedance-based frequency scans to assess the SSCI risk of large data center loads. In a detailed proprietary model, the motor, VFD, and lighting branches stayed passive from 5 to 1000~Hz, while the UPS-fed IT branch in double-conversion mode showed negative resistance across the whole sub-synchronous band. Three open-source models built by other groups confirmed the pattern: no negative resistance in bypass mode, and negative resistance only in double-conversion mode, with a size and frequency range that depend on the converter controls. In every model where it appeared, a lighter load increased the magnitude of the negative resistance. So a lightly loaded facility in double-conversion mode is the worst case in terms of negative resistance. Raising the DC-link capacitance from 20 to 50 mF confined the negative resistance of the INS-ISU model to a narrower band around the fundamental, consistent with a reduction in the closed-loop bandwidth of the DC-voltage controls. This result indicates that the DC-voltage control response is a key contributing factor to the negative resistance characteristic of the double conversion UPS. For interconnection studies, we recommend screening the UPS branch in double-conversion mode at partial load and, where negative resistance appears, running a combined scan with the network at the POI. Future work will cover combined load--network scans for series-compensated and weak grids, EMT time-domain confirmation of the predicted oscillations, and hardware or field validation.

\appendix
\section{Further Exploration on Open-source Models}

\begin{figure}[!htbp]
  \centering
\begin{minipage}[t]{0.485\linewidth}
  \centering
  \includegraphics[width=\linewidth]{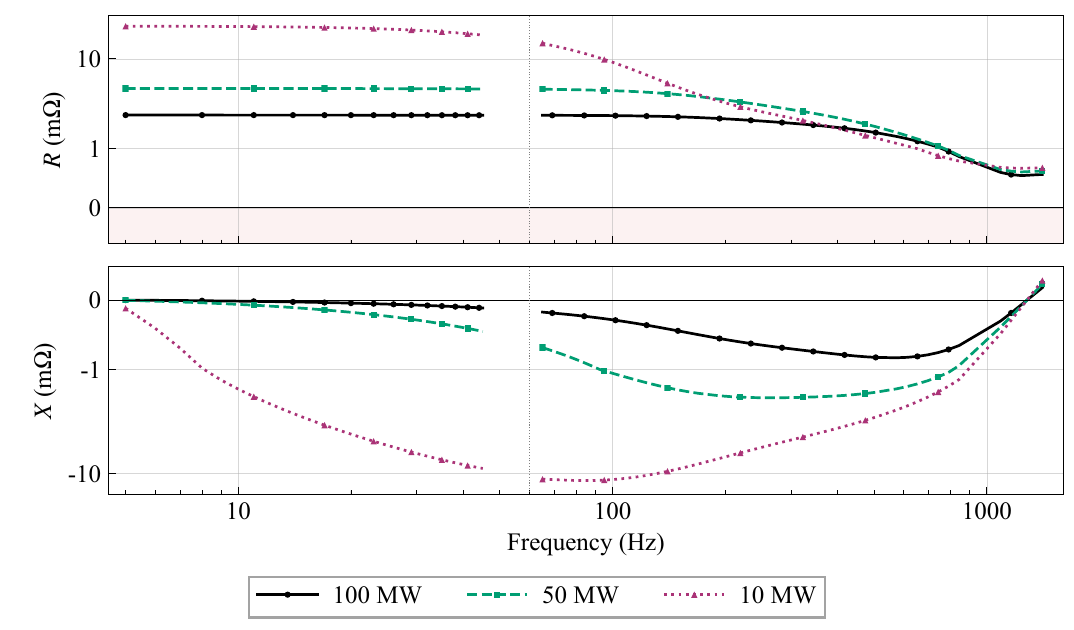}
  \caption{Positive-sequence impedance of the PNNL model in bypass mode at three loading levels. The shaded area marks negative resistance.}
  \label{pnnl_bypass_load}
\end{minipage}\hfill
\begin{minipage}[t]{0.485\linewidth}
  \centering
  \includegraphics[width=\linewidth]{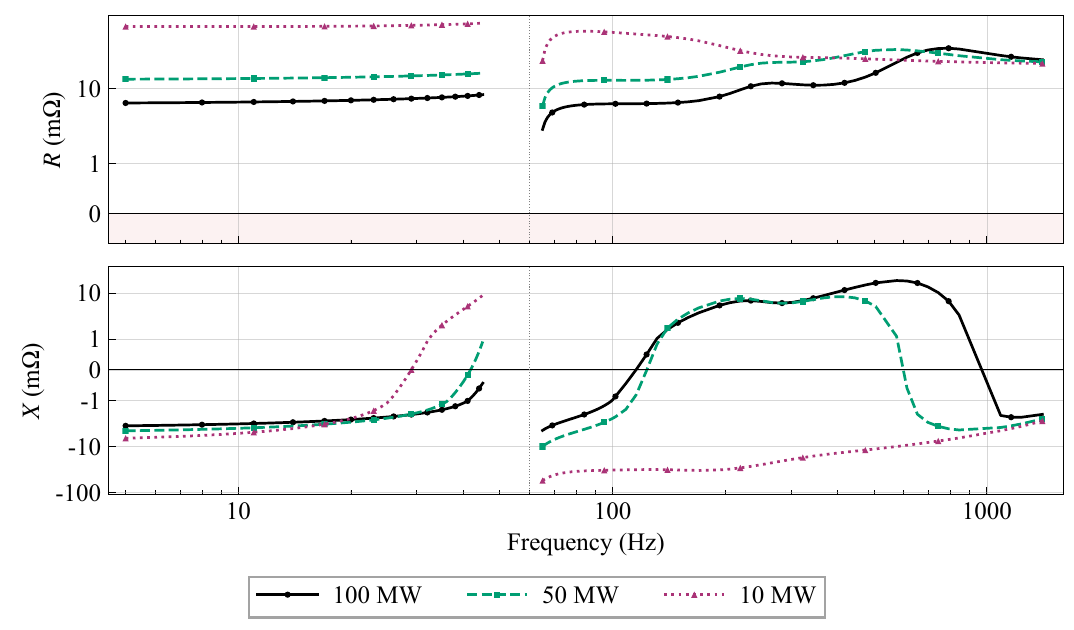}
  \caption{Positive-sequence impedance of the PNNL model in double-conversion mode at three loading levels. The shaded area marks negative resistance.}
  \label{pnnl_dcm_load}
\end{minipage}
\end{figure}

\begin{figure}[!htbp]
  \centering
\begin{minipage}[t]{0.485\linewidth}
  \centering
  \includegraphics[width=\linewidth]{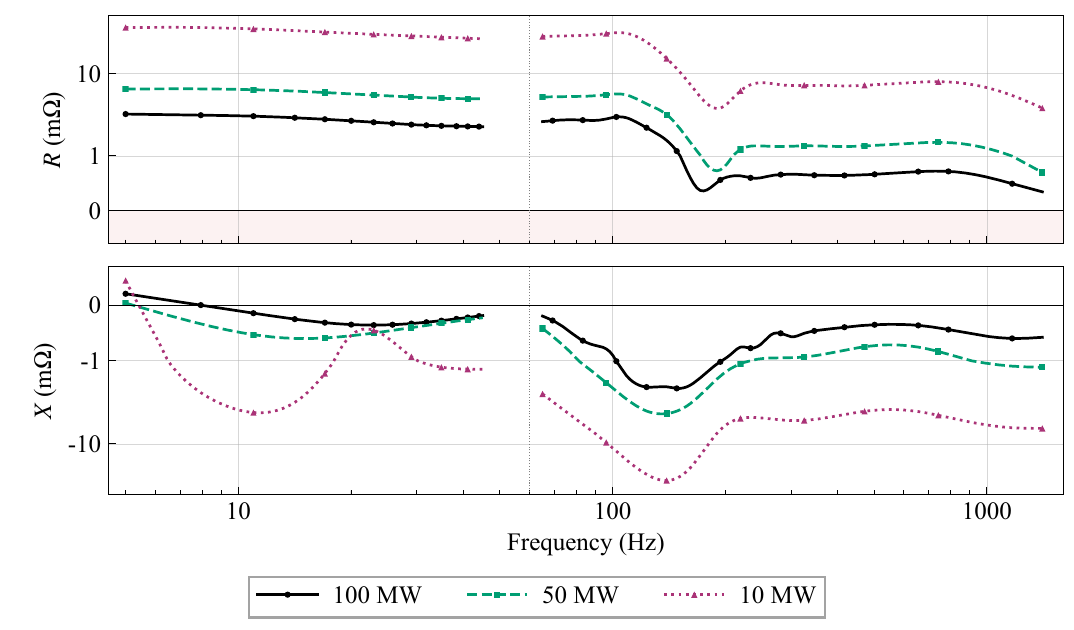}
  \caption{Positive-sequence impedance of the TAMU model in bypass mode at three loading levels. The shaded area marks negative resistance.}
  \label{tamu_bypass_load}
\end{minipage}\hfill
\begin{minipage}[t]{0.485\linewidth}
  \centering
  \includegraphics[width=\linewidth]{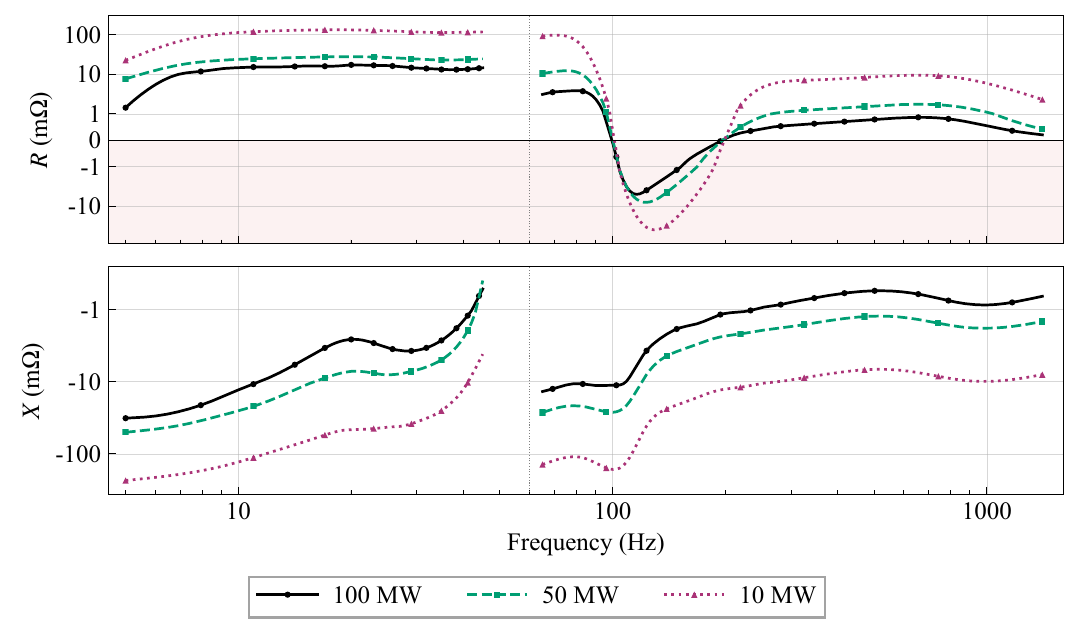}
  \caption{Positive-sequence impedance of the TAMU model in double-conversion mode at three loading levels. The shaded area marks negative resistance.}
  \label{tamu_dcm_load}
\end{minipage}
\end{figure}

\begin{figure}[!htbp]
  \centering
\begin{minipage}[t]{0.485\linewidth}
  \centering
  \includegraphics[width=\linewidth]{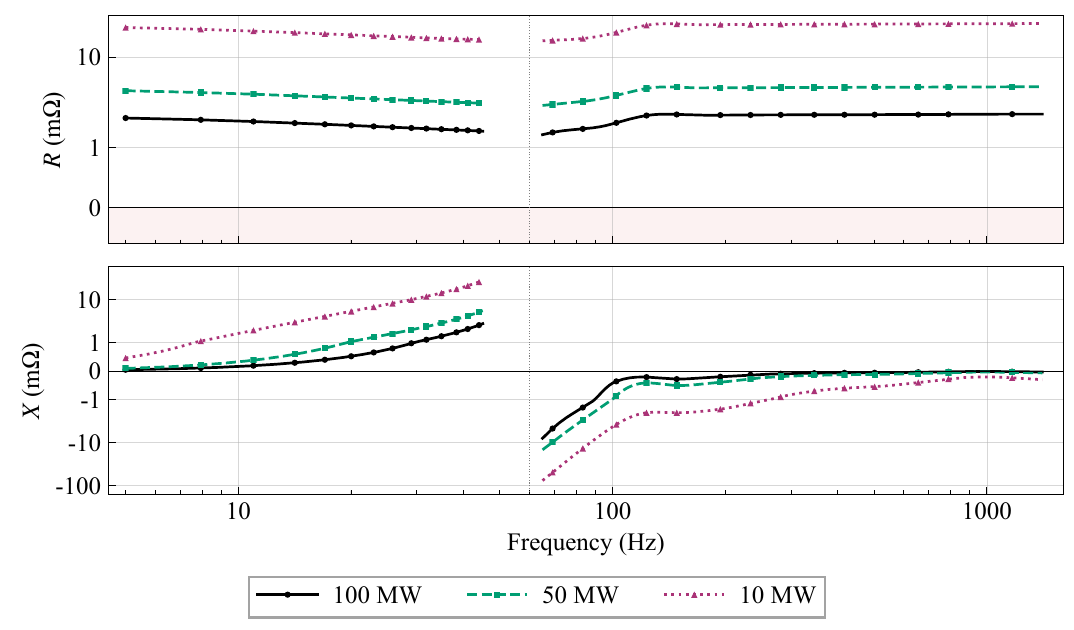}
  \caption{Positive-sequence impedance of the NLR model in bypass mode at three loading levels. The shaded area marks negative resistance.}
  \label{nrel_bypass_load}
\end{minipage}\hfill
\begin{minipage}[t]{0.485\linewidth}
  \centering
  \includegraphics[width=\linewidth]{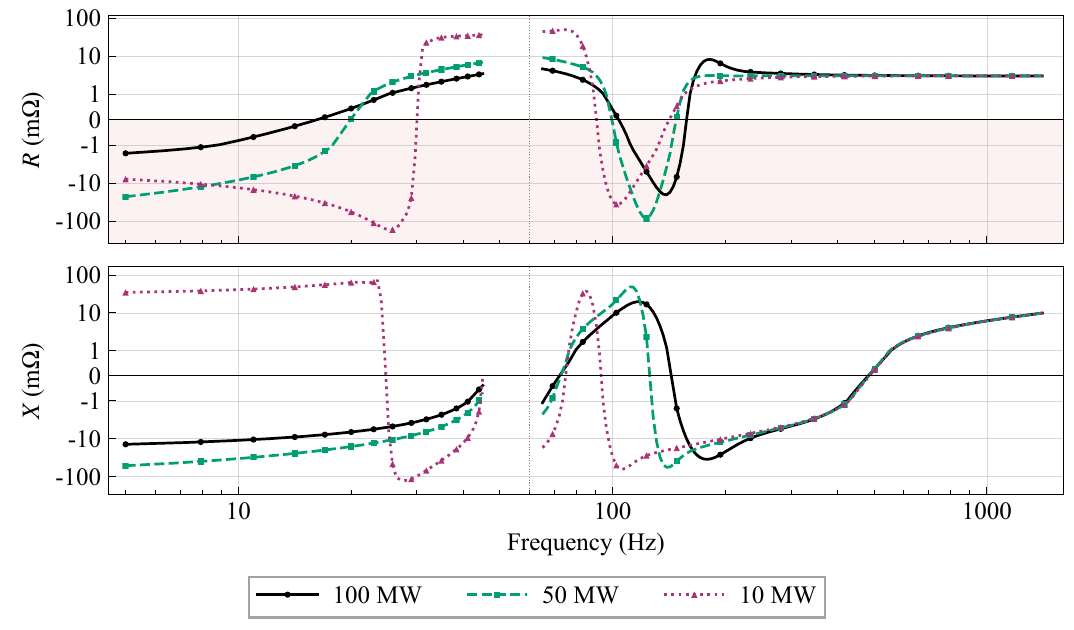}
  \caption{Positive-sequence impedance of the NLR model in double-conversion mode at three loading levels. The shaded area marks negative resistance.}
  \label{nrel_dcm_load}
\end{minipage}
\end{figure}

Figs.~\ref{pnnl_bypass_load}--\ref{nrel_dcm_load} repeat the loading study of Section~\ref{sec:sensitivity} for the three open-source models at 100, 50, and 10~MW. In bypass mode (Figs.~\ref{pnnl_bypass_load}, \ref{tamu_bypass_load}, and \ref{nrel_bypass_load}), all three models keep a positive resistance over the full range, and it grows roughly in inverse proportion to the load: about 1.5--3~m$\Omega$ at 100~MW and about 15--30~m$\Omega$ at 10~MW below the fundamental. The smallest margin, about 0.4~m$\Omega$, is in the TAMU model near 170~Hz at 100~MW. This agrees with the INS-ISU bypass results in Fig.~\ref{isu_bypass_load}. In double-conversion mode, the PNNL model (Fig.~\ref{pnnl_dcm_load}) stays positive at all loads, at about 6, 13, and 50~m$\Omega$ below the fundamental at 100, 50, and 10~MW. The TAMU model (Fig.~\ref{tamu_dcm_load}) is positive below the fundamental but has a negative band between about 100 and 190~Hz, whose minimum deepens from about $-5$~m$\Omega$ at 100~MW to about $-25$~m$\Omega$ at 10~MW. The NLR model (Fig.~\ref{nrel_dcm_load}) depends most on loading. Its low-frequency negative band extends up to about 17~Hz at 100~MW, 22~Hz at 50~MW, and 33~Hz at 10~MW, where it reaches about $-150$~m$\Omega$, and its band between about 90 and 160~Hz reaches $-20$ to $-100$~m$\Omega$. The models differ in where and how much negative resistance they show, but none goes negative in bypass mode, and where double-conversion mode produces negative resistance, a lighter load generally makes it deeper or wider. The open-source models therefore support the conclusion of Section~\ref{sec:sensitivity} that a lightly loaded data center in double-conversion mode is the worst case for SSCI.

\end{document}